\documentclass[prl,amsmath,amssymb]{revtex4-1}

\usepackage{graphics}
\usepackage{graphicx}
\usepackage{amsmath}
\usepackage{amssymb}
\usepackage{xcolor}
\usepackage{tikz}
\usepackage{siunitx}
\usepackage{booktabs} % For prettier tables
\usepackage{multirow} % Two merge multiple lines
\usepackage{bigdelim}% Giant bracket in table
\usepackage{amssymb,pifont}% Special symbols (empty star and diamond)
\usepackage{derivative}% pour utiliser le symbole de dérivée partielle
\usepackage{lipsum}% pour mettre du texte phantome 
\usepackage[T1]{fontenc}
\usepackage{newtxtext}
\usepackage{newtxmath}
\usepackage{hyperref}
\hypersetup{
    colorlinks = true,
    urlcolor   = blue,
    citecolor  = black,
}

\definecolor{f2c1}{rgb}{0.792, 0.871, 0.941}
\definecolor{f2c2}{rgb}{0.376, 0.655, 0.824}
\definecolor{f2c3}{rgb}{0.153, 0.467, 0.722}
\definecolor{f2c4}{rgb}{0.031, 0.302, 0.588}
\definecolor{f2c5}{rgb}{0.031, 0.216, 0.463}

\definecolor{f3c1}{rgb}{0.792156862745098,0.8705882352941177,0.9411764705882353}
\definecolor{f3c2}{rgb}{0.3764705882352941,0.6549019607843137,0.8235294117647058}

\newcommand{\C}[1]{\textcolor{black}{#1}}

\usepackage{tikz}
\begin{document}
\title{The macroscopic contact angle of water on ice}

\author{W. Sarlin, D.V. Papa, R. Grivet, A. Rosenbaum}
\affiliation{Laboratoire d'Hydrodynamique, CNRS, Ecole polytechnique, Institut Polytechnique de Paris, Palaiseau, 91120, France}
\author{A. Huerre}
\affiliation{Laboratoire Mati\`ere et Syst\`emes Complexes (MSC), Universit\'e Paris Cit\'e, CNRS, UMR 7057, Paris, 75013, France}
\author{T. S\'eon}
\affiliation{Institut Franco-Argentin de Dynamique des Fluides pour l'Environnement (IFADyFE), CNRS (IRL 2027), Universidad de Buenos Aires, CONICET, Buenos Aires, 1428, Argentina}
 \author{C. Josserand}
 \affiliation{Laboratoire d'Hydrodynamique, CNRS, Ecole polytechnique, Institut Polytechnique de Paris, Palaiseau, 91120, France}
%\corresau{Wladimir Sarlin, \email{wladimirsarlin@outlook.com}}

\begin{abstract}
Wettability quantifies the affinity of a liquid over a substrate, and determines whether the surface is repellent or not. When both the liquid and the solid phases are made of the same chemical substance and are at thermal equilibrium, complete wetting is expected in principle, as observed for instance with drops of molten metals spreading on their solid counterparts. However, this is not the case for water on ice. Although there is a growing consensus on the partial wetting of water on ice and several estimates available for the value of the associated contact angle, the question of whether these values correspond to the equilibrium angle without thermal effects is still open. 
In the present paper, we address this issue experimentally and demonstrate the existence of a macroscopic contact angle of water on ice using theoretical arguments. 
Indeed, when depositing water droplets on smooth ice layers with accurately controlled surface temperatures, we observe that spreading is unaffected by thermal effects and phase change close enough to the melting point. %, where the liquid film relaxes to a stable sessile drop shape.
Whereas the short time \C{motion of the contact line} is driven by an inertial-capillary balance, the evolution towards equilibrium is described by a viscous-capillary dynamics and is therefore capillary - and not thermally - related. Moreover, we show that this contact angle remains constant for undercoolings below 1 K. 
This way, we show the existence of a non-zero equilibrium contact angle of water on ice, that it is very close to 12$^\circ$.
%the measured macroscopic contact angle at the end of the spreading process is \C{representative of the equilibrium one}. 
We anticipate this key finding to significantly improve the understanding of capillary flows in the presence of phase change, which is especially useful in the context of ice morphogenesis and of glaciology, but also in the aim of developing numerical methods for resolving triple-line dynamics.
%Furthermore, a progressive decrease of the contact angle towards this equilibrium angle for ice temperature approaching the melting point is observed.
\end{abstract}
\maketitle

\section{Introduction}
\label{introduction}

Wetting refers to the science of how a liquid deposited on a solid (or liquid) substrate spreads out. It plays an important role in many aspects of our everyday life such as the first breath of a newborn, underground flows, painting, chemistry, automobile, food industry, 3D printing, sap flow in plants or even eyes lubrication \citep{1985_de_gennes,2009_bonn}. A case of particular interest is that of the wetting of water on its own solid phase, ice. Indeed, a wide variety of ice structures are formed by the freezing of capillary flows (such as drops, rivulets and liquid films), including ice accretion on airplanes \citep{2001_lynch}, power lines \citep{1998_laforte}, bridge cables \citep{2019_liu} or wind turbines \citep{2017_wang}; ice falls \citep{2010_montagnat}; icicles \citep{2011_chen}; cloud physics \citep{2006_dash}; frost heave \citep{2006_wettlaufer}; frozen rivers \citep{2013_beltaos}; and aufeis \citep{1990_schohl}. However, despite this ubiquitous coexistence of ice and capillary flows, the wetting behavior of water on ice remains largely unknown.

%%% Angle micro :
An important piece of experimental developments in the last decade has been devoted to the characterization of the shape of a water-like layer at the ice surface \citep{2012_sazaki, 2016_asakawa, 2019_slater}. Notably, these authors reported that as ice premelts a wetting quasi-liquid layer forms on the ice and thickens up to a nanometric thickness that depends on temperature \citep{2016_murata}, before the film free-surface gets attracted by the ice. This leads to the so-called pseudopartial wetting situation \citep{1991_brochard-wyart}, where the quasi-liquid dewets and form micrometric water droplets that stand on the film.

The equilibrium contact angle, $\theta_e$, that a water drop makes with the ice at thermodynamic equilibrium can be expressed as a function of the three surface tensions $\gamma_{iw}$, $\gamma$ and $\gamma_{iv}$ associated with the solid-liquid, liquid-vapour and solid-vapour interfaces, respectively, using Young-Dupr\'e relation

\begin{equation}
\gamma_{iv} = \gamma_{iw} + \gamma\cos\theta_e.
\label{eq:young}
\end{equation}

%%% Valeur angle de contact, une grande incertitude dans la valeur de l'angle macro :
At the microscopic scale, the contact angle of micrometric quasi-liquid water droplets has been measured to be between $0.6^\circ$ and $2.3^\circ$ \citep{2016_murata}, while a theoretical analysis predicted a value of $3.4^\circ$ \citep{2022_luengo-marquez}. 
By contrast, at macroscopic scale there is a much larger discrepancy in the reported values for the water-ice contact angle \citep{2025_huerre}. The first attempt to measure this quantity was proposed 60 years ago by \citet{1967_knight}, who deposited a puddle of hot water on a cold copper substrate and measured a receding contact angle of 12$^\circ$ at a temperature slightly below 0$^\circ$C. A few years later, the same author reported other measurements in various configurations \citep{1971_knight} and found different values, before ultimately concluding that measuring this angle unequivocally may be impossible, although he thought that the complete wetting scenario is very unlikely \citep{1996_knight}. Since this seminal work, other authors conducted similar experiments, depositing either a water drop or film on ice and measuring the advancing or receding contact angle, and found values spanning a large range from 6$^\circ$ up to 40$^\circ$ \citep{1997_makkonen, 2020a_thievenaz, 2023_demmenie, 2024_grivet}. In a very recent study, \citet{2025_demmenie} performed droplet deposition experiments involving water and ice with varying surface temperature. These authors showed how the apparent angle varies with temperature and also identify 12$^\circ$ as a possible value for the contact angle of water on ice. However, they did not approach the melting point accurately, so that they could not demonstrate that this angle is reached in the absence of thermal effects.

%Nevertheless, this value was obtained using a semi-empirical model involving a fitting parameter and without approaching accurately enough the melting point.

%however this study mainly focused on cold substrate temperature. The lack of precision near the melting point and the semi-empirical model they propose, relying on a fitting parameter without clear physical basis, prevented these authors from bringing a definitive conclusion.}

%%% Equilibre thermodynamique et valeur de la tension de surface de ice-vapour
This significant discrepancy of the results is probably due to the fact that all these experiments have been realised far from thermodynamic equilibrium, precisely because reaching such a limit remains an experimental challenge. 
A possible path to know the contact angle of water on ice at thermodynamic equilibrium would be to deduce this angle by considering the Young-Dupr\'e relation. Indeed, given that the surface tension of water at 0 $^\circ$C is $\gamma$ = 75.6 mJ.m$^{-2}$ \citep{2004_lide}, $\theta_e$ could in theory be estimated using relation \eqref{eq:young}, if both $\gamma_{iv}$ and $\gamma_{iw}$ are available. Unfortunately, these values are not known precisely enough today to discriminate between 0 and $50^\circ$, as $\gamma_{iw}$ is currently found to be in the range 27-35 mJ.m$^{-2}$ \citep{2016_espinosa, 2017_ambler} and $\gamma_{iv}$ in the range 70-120 mJ.m$^{-2}$ \citep{1957_de_reuck, 1969_ketcham, 1992_van_oss, 2004_pruppacher, 2017_djikaev} at 0$^\circ$C.

%\cite{Makkonen1997} proposed that static equilibrium contact angle of a water drop at temperature close to 95$^\circ$C deposited on ice at -25$^\circ$C is around 37$^\circ$.
%Recently, \cite{demmenie_growth_2023} conducted experiments of sessile water droplets on ice and concluded that the contact angle varies with the ice temperature from 40$^\circ$ at -10$^\circ$C up to 12$^\circ$ close to the melting temperature.

%In the experiment of \cite{thievenaz_retraction_2020} of drop impact on a cold surface described at the end of section \ref{SpreadCarac}, we saw that, after a given time, a film of water retracts on a growing ice layer and forms either a drop or a ring (resp. Fig.~\ref{fig:splat}a and b)
%In the first case, when the retracting water film has time to build an equilibrium spherical cap before being frozen, the authors show that its contact angle with the ice always reaches a constant limit value of $\sim 12^\circ$, for any control parameters. This angle is an equilibrium or a receding angle, but in either case, it exists, is unique and constant.
%Finally, recently \cite{Grivet2024} carried out a Landau-Levich-type experiment in which an ice sheet is pulled out of a water bath. At low speeds, there is a range of parameters in which the water film is not entrained, and its height along the ice allows the authors to deduce a receding contact angle between water and ice. This original measurement of the contact angle of water on ice predicts a value of 6$^\circ$ with an accuracy of a few degrees.

\begin{figure*}[t]
  \includegraphics[width=\linewidth]{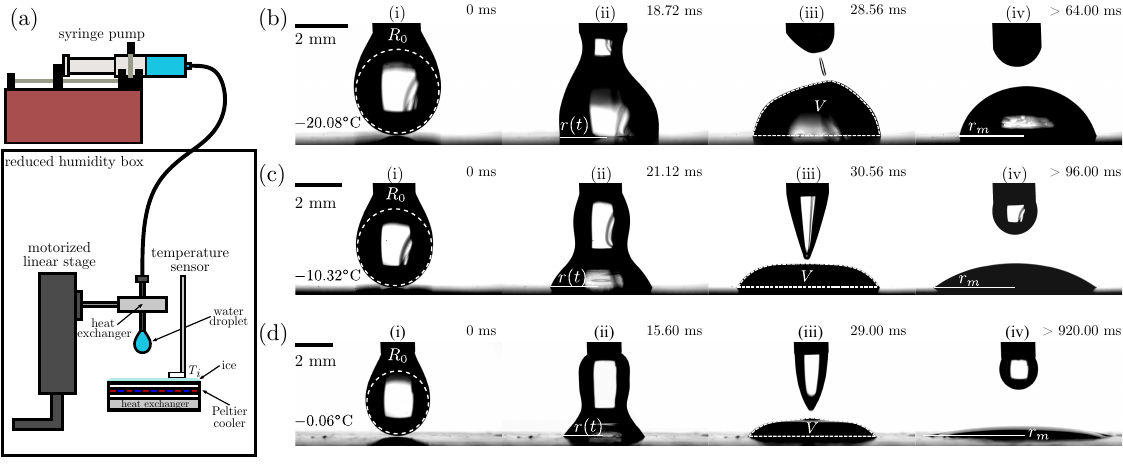}% Here is how to import EPS art
  \caption{\label{fig:setup} (a) Schematics of the experimental setup. (b) \C{Image sequence of the} spreading of a \C{water} droplet on ice at a low \C{surface} temperature ($T_i = -20.08$ \textcelsius). \C{Four representative stages are displayed}: (i) Just before \C{the initial} contact with the \C{ice} surface ($t=0~$ms), (ii) at short time ($t=18.72~$ms), (iii) when the drop detaches from the needle ($t=29.76~$ms) and (iv) when motion stops but before complete freezing  ($t > 64.00~$ms). (c)-(d) Spreading of a droplet on ice at (c) a moderately low surface temperature ($T_i = -10.32$ \textcelsius) and (d) \C{close to the melting point of water} ($T_i = -0.06$ \textcelsius). The same four stages (i)-(iv) are depicted. The initial radius of curvature $R_0$ of the pendent drop, the radius $r(t)$ of the wetted area, the drop volume $V$ and the arrest radius $r_m$ are highlighted in white.}
\end{figure*}

%%% Conclusion and motivation 
%Therefore, even thought the existence of a non-zero contact angle between water and ice is well accepted nowadays, there is absolutely no consensus yet on its value and on the surface tension of the ice-air interface \citep{2025_huerre}. 

In the present study, we perform experiments in which a water droplet is deposited on a smooth ice layer whose surface temperature is meticulously varied, in order to approach thermal equilibrium (\textit{i.e.}, temperature equilibrium between solid and liquid phases) as closely as possible. For temperatures near the \C{melting} point, we demonstrate that contact line spreading and arrest do occur independently of thermal effects and phase change.
Under these near-thermal equilibrium conditions, we experimentally observe a unique contact angle of water on ice that is very close to the equilibrium contact angle obtained in the thermodynamic limit.
%mettre ailleurs je pense !
%Given that small hysteresis of the contact angle is expected in this case, because the ice surface is very  smooth and dry, this measured angle can be considered to be very close to the equilibrium contact angle obtained in the thermodynamic limit}. 
This allows us to report its value and deduce a fine estimate of the interfacial tension $\gamma_{iv}$ between ice and air at the triple point.
Furthermore, we characterize the evolution of the contact angle of water on ice with temperature out of thermodynamic equilibrium.

\section{Experimental setup and qualitative results}
\label{setup}

A schematic of the experimental setup is illustrated in figure \ref{fig:setup}(a). Experiments consist in depositing a pendent drop of pure demineralized water of density $\rho$, viscosity $\eta$ and surface tension $\gamma$ at vanishing impact velocity (using a linear stage) on a flat layer of ice made of the same pure demineralized water, whose surface temperature is $T_i$. The spreading of the droplet is recorded with a Photron FASTCAM SA-X2 high-speed camera operating between 8000 and 12500 frames per second. Both drop and air temperatures ($T_d$ and $T_a$, respectively) are controlled and kept as close as possible to the melting temperature of water, \C{$T_f = 273.15$ K}. Experiments are carried out in a reduced relative humidity environment ($RH_a \simeq 16 \pm 2$ \%) in order to limit the effect of this parameter on the ice and frost formation \citep{2021_sebilleau}. Each test is performed at a constant $T_i$ and is repeated five times to ensure reproducibility of the results. In the present study, $T_i$ is varied from close to the melting point up to far out of equilibrium, \textit{i.e.}, $T_i \in $ [$-0.03$ \textcelsius, $ -20.42$ \textcelsius], while all other parameters ($T_d$, $T_a$, $RH_a$) are kept constant.
 
The ice surface is formed by freezing a small layer of water of thickness $\sim 5$ mm with a 40$\ \times\ $40 mm Peltier cooler mounted on a heat exchanger and connected to a recirculating thermal bath that operates at $T_\mathrm{bath} = 1 \pm 0.1$ \textcelsius. The bath ensures a proper heat evacuation from the Peltier cooler and therefore a precise control of the thermal flux applied when imposing a voltage. A smooth and mirror-like ice surface is obtained by enforcing a low thermal flux on the Peltier modulus, typically of the order of $0.1$ K/min, that leads to a slow and controlled growth of the solidification front. The surface temperature of the ice layer, $T_i$, is measured by placing a class A flat PT100 RTD sensor (precision of 0.01 \textcelsius) on top of the ice surface and very close to the deposition site of the droplet (see figure \ref{fig:setup}(a)). The thermal flux is then gently adapted to set the desired ice temperature $T_i$. The drop temperature, $T_d$, is controlled by passing the needle where the pendent drop is formed through a heat exchanger also connected to the recirculating bath. $T_d$ is monitored with a type K thermoelectrical sensor located inside the tip of the needle. Throughout the study, its value was kept at $T_d = 2.5 \pm 0.3$ \textcelsius. The air temperature $T_a$ is controlled by injecting cold nitrogen gas in the box and is measured close to the deposition site with a type K thermoelectric sensor. This parameter is roughly kept constant at $T_{a} = 7 \pm 1.5$ \textcelsius. The same cold nitrogen influx allows us to set the value of $RH_a$, that is controlled using a domestic humidity sensor. \C{With such a small value for the reduced relative humidity, we obtained an ice surface which is almost completely dry, in addition of being very smooth.} %Both $T_d$ and $T_a$ are kept as close as possible to the melting point of water $T_f$.

The video recordings are used to monitor the spreading dynamics and to determine all the variables of interest. The initial radius of curvature $R_0$ of the pendent drop and the water-air surface tension $\gamma$ are obtained from the last image preceding the onset of spreading (column (i) in figures \ref{fig:setup}(b)-(d)) using the pendent drop method introduced by \citet{2016_daerr}. The radius $r(t)$ of the wetted area is extracted from the contour of the spreading droplets (column (ii) in figures \ref{fig:setup}(b)-(d)). The volume $V$ of the drop is computed from the first image following the droplet's detachment from the tip of the needle, when the contrast in the images is at its maximum (column (iii) in figures \ref{fig:setup}(c)-(d)). For cold ice surfaces whose undercooling $\Delta T = T_f - T_i \geqslant 11\ \rm{K}$, the drop does not detach from the needle axisymmetrically, hence $V$ is not computed for these tests. % as in \ref{fig:setup} (b). This point will be discussed in further details later on. 
At the end of the spreading process, an extra image is taken (column (iv) in figures \ref{fig:setup}(b)-(d)) to compute the arrest radius $r_m$ of the drop and the apparent contact angle $\theta_a$ that the droplet forms with the ice. $\theta_a$ is determined using an adapted version of the algorithm introduced by \citet{2020_quetzeri-santiago}. Furthermore, the density and viscosity of water are taken as $\rho = 1000 \ \mathrm{kg.m^{-3}}$ and $\eta = 1.8 \times 10^{-3}\ \rm{Pa.s}$, respectively.

%both evaluated at the drop temperature $T_d = 2.5$ \textcelsius \ and

The spreading of the droplet is illustrated in figure \ref{fig:setup} for three different surface temperatures $T_i = -20.08$ \textcelsius, $T_i = -10.32$ \textcelsius\ and $T_i = -0.06$ \textcelsius\ (panels (b), (c) and (d), respectively) and at four characteristic moments (columns (i)-(iv)). At the beginning of an experiment, a pendent drop of initial radius of curvature $R_0$ is put into contact with the ice surface (i) on which it starts spreading. The spreading rate is quantified by considering the radius of the wetted area of the drop on the ice surface, $r(t)$ (ii). At some point in the dynamics, a volume $V$ of liquid detaches from the needle (iii). When spreading ceases, this volume eventually stabilizes to form a sessile droplet which exhibits a spherical cap shape of arrest radius $r_m$ (iv). For large undercoolings $\Delta T$, the drop shape is similar to a hemisphere and its apparent contact angle $\theta_a$ with the substrate is high (figure \ref{fig:setup}(b)(iv)), whereas when the ice surface temperature approaches the melting point of water $T_f$ the sessile drop resembles a very thin lens with a small contact angle (figure \ref{fig:setup}(d)(iv)). 

\section{Results and discussion}
\label{res}

\subsection{Spreading dynamics}
\label{spreading}

\begin{figure*}[t]
\includegraphics[width=\linewidth]{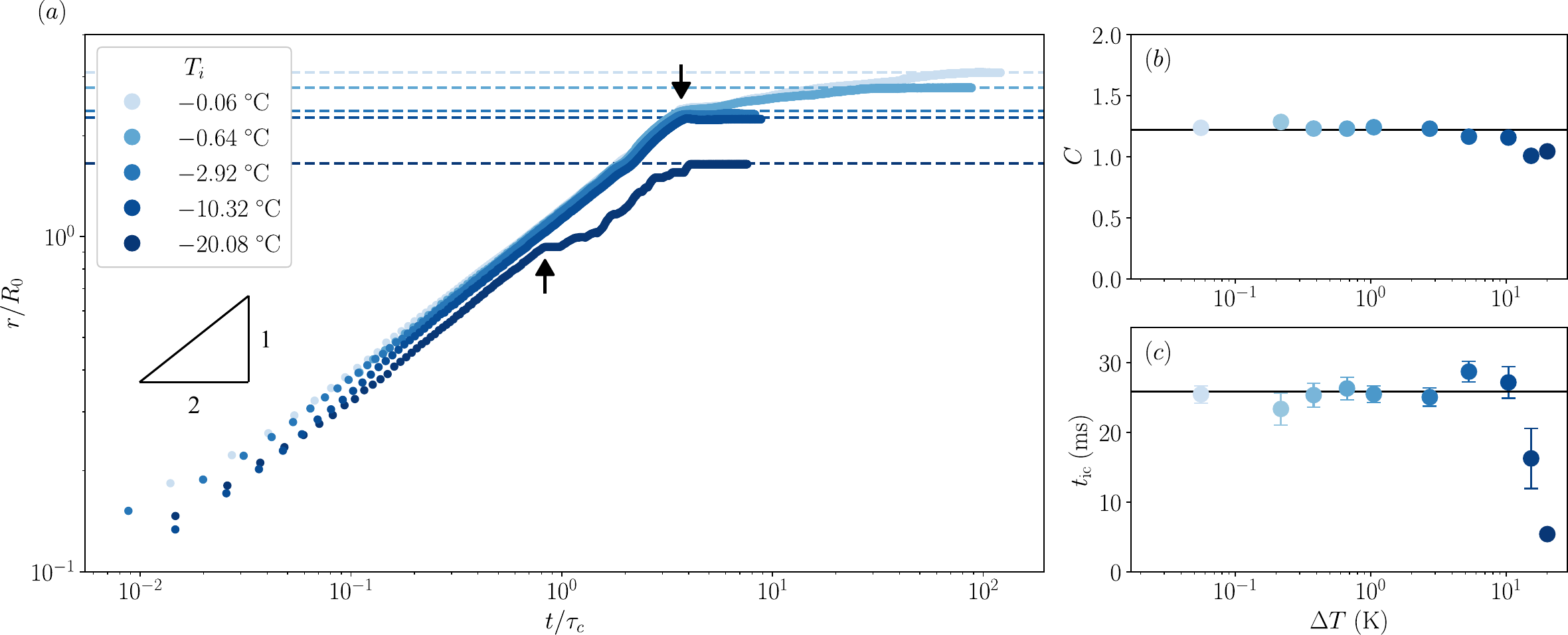}% Here is how to import EPS art
\caption{\label{fig:radius} (a) Dimensionless spreading radius, $r/R_0$, as a function of the rescaled time, $t/\tau_c$, in a logarithmic scale. The horizontal dashed lines highlight the arrest radius reached for each experiment. (b) Evolution of the numerical prefactor $C$ from equation \eqref{inertio_capillary_scaling} with the undercooling $\Delta T = T_f - T_i$, with $T_f$ the melting point of water. The solid line represents $C=1.22$. (c) $t_{\rm{ic}}$ as a function of $\Delta T$. The solid line indicates $t_{\rm{ic}} = 25.84\ \rm{ms}$. The colors of the symbols denote the surface temperature $T_i$ of the ice, with darker shades corresponding to colder ice.}
\end{figure*}

The evolution of the dimensionless radius $r/R_0$ with the rescaled time $t/\tau_c$ is presented in figure \ref{fig:radius}(a) for five representative ice surface temperatures, $T_i$. Here, $\tau_c = \sqrt{\rho {R_0}^3/\gamma}$ is the inertial-capillary timescale. A first phase of fast spreading is observed for all experiments (before the black arrows), which is reminiscent of the short-time evolution observed for isothermal droplet deposition experiments \citep{2004_biance,2008_bird,2012_winkels}. This regime ends at a time $t_{\rm{ic}}$ which depends on the ice temperature. For the dark blue curve in figure \ref{fig:radius}(a), corresponding to $T_i = -20.08$ \textcelsius\ (\textcolor{f2c5}{$\bullet$}) and to sequence (b) in figure \ref{fig:setup}, a partial-pinning of the contact line is observed from $t/\tau_c \simeq 0.8$ up to $t/\tau_c \simeq 4.1$, when the arrest radius is reached. Such a behavior is illustrated by the asymmetry of the free-surface of the spreading drop in figure \ref{fig:setup}(b)(ii) (compared to figures \ref{fig:setup}(c)(ii) or \ref{fig:setup}(d)(ii)). This observation suggests that, for cold enough ice surface, the contact line can undergo pinning and unpinning due to interactions between the advancing contact line and the ice layer that grows beneath the spreading drop. For moderate ice surface temperatures, a different evolution is noticed, as illustrated by the two curves corresponding to $T_i = -2.92$ \textcelsius\ (\textcolor{f2c3}{$\bullet$}) and $T_i = -10.30$ \textcelsius\ (\textcolor{f2c4}{$\bullet$}) in figure \ref{fig:radius}(a). Indeed, for these experiments the fast spreading dynamics abruptly stops around $t/\tau_c \simeq 3.8$ without any significant \textcolor{black}{partial-pinning} of the contact line afterwards. Finally, for ice surface temperatures closer to the melting point, a second and slower dynamics takes place after the sudden stop of the initial fast spreading regime. This is illustrated by the light blue curves in figure \ref{fig:radius}(a), corresponding to $T_i = -0.06$ \textcelsius \ (\textcolor{f2c1}{$\bullet$}) and to $T_i = -0.64$ \textcelsius \ (\textcolor{f2c2}{$\bullet$}),  from $t/\tau_c \simeq 3.8$ onward. %from $t_{\rm{ic}}/\tau_c \simeq 5.7$ onward. 
For all experiments, the horizontal dashed line whose color corresponds to the symbols indicates the arrest radius $r_m$.

%For all experiments, the horizontal dashed line with the same color as the symbols indicate the arrest radius, determined from a picture taken long after the spreading process has stopped.

%\subsection{The inertial-capillary spreading}
%\label{ic_reg}

The short-time dynamics is well described by a balance between Laplace pressure, that drives the motion, and the inertial pressure that resists it \citep{2004_biance,2008_bird,2012_winkels,2022_grivet}. Indeed, equating the two effects gives the following inertial-capillary dynamics for the radius of the wetted area 

\begin{equation}
    \label{inertio_capillary_scaling}
    \frac{r(t)}{R_0} = C \left( \frac{t}{\tau_c} \right)^{1/2},
\end{equation}

%(a) $t_{\rm{ic}}$ as a function of $\Delta T$. The solid line represents $t_{\rm{ic}} = 25.84\ \rm{ms}$. (b)

%\C{(see details in Supplemental Material)}

\noindent with $C$ a numerical prefactor. It can be observed in figure \ref{fig:radius}(a) that the short-time evolution of the dimensionless radius, before each of the black arrows, effectively displays a square root of time evolution, as evidenced by the 1/2 slope followed by all curves in this first dynamical regime. \C{It has been verified that this $1/2$ power law is insensitive to errors in determining the time origin, given the large temporal resolution at which the experiments were recorded}. The prefactor $C$ is systematically extracted and presented in figure \ref{fig:radius}(b) as a function of the undercooling $\Delta T = T_f - T_i$. For $\Delta T < 11\ \rm{K}$, $C$ plateaus at a constant value of 1.22, \C{that is consistent with the measurements reported by \citet{2008_bird} ($C \in [0.75, 1.5]$) and \citet{2022_grivet} ($C = 1.15 \pm 0.06$) for droplet deposition experiments}. For undercoolings beyond 11 K, a slight decrease of $C$ is noticed, that can be attributed to the increasing rate of solidification. Strikingly, the constant value of $C$ over two decades of $\Delta T$ between 0.06 and 11 K reveals that this inertial-capillary regime is unaffected by the ice surface temperature. In other words, in this regime the contact line dynamics of the spreading droplet is insensitive to both the temperature gradients and the phase change at short times. The arrest time of this inertial-capillary dynamics, $t_{\rm{ic}}$, is then determined for all experiments. Its evolution with the undercooling $\Delta T$ is displayed in figure \ref{fig:radius}(c), where one may note that $t_{\rm{ic}}$ is constant when $\Delta T < 11\ \rm{K}$ with $t_{\rm{ic}} \simeq 25.84\ \rm{ms}$, but drops sharply for larger undercooling. This demonstrates that, for $\Delta T > 11\ \rm{K}$, solidification does induce an early pinning of the advancing contact line during the inertial-capillary regime, whereas for smaller undercooling the arrest of this dynamics is not thermally-related. 

%\subsection{The viscous-capillary regime}
%\label{cv_reg}

Then, two different situations can occur when $\Delta T < 11\mathrm{K}$. In the intermediate range $1 \ \rm{K} < \Delta T < 11\ \rm{K}$, the spreading of the droplet ceases right after the inertial-capillary regime, as highlighted by the experiments at $T_i = -2.92$ \textcelsius\ (\textcolor{f2c3}{$\bullet$}) and $T_i = -10.3$ \textcelsius\ (\textcolor{f2c4}{$\bullet$}) in figure \ref{fig:radius}(a), where no motion is observed after $t_{\rm{ic}}$. However, a second and slower dynamics systematically takes place when $\Delta T < 1\mathrm{K}$, as illustrated by experiments where $T_i = -0.06$ \textcelsius \ (\textcolor{f2c1}{$\bullet$}) and $T_i = -0.64$ \textcelsius\ in figure \ref{fig:radius}(a). We suspect this second regime to correspond to a viscous-capillary spreading: Indeed, without thermal effects such a dynamics usually develops once inertia becomes negligible \citep{2004_biance,2009_bonn}. As discussed in particular by \citet{1984_de_gennes}, this behavior is intimately related to wettability that governs the manner the contact line relaxes to equilibrium. In case of total wetting, one would expect Tanner's law to be valid, so that $r/R_v \propto (\gamma t/(\eta R_v))^{1/10}$, with $R_v=(3V/(4\pi))^{1/3}$ the equivalent radius for a droplet of initial volume $V$ \citep{1979_tanner,1984_de_gennes}. However, if partial wetting is at play, a \C{viscous-capillary spreading similar to a Cox-Voinov dynamics} is expected for the moving contact line \citep{1976_voinov,1986_cox,2009_bonn}. \C{To describe the motion of the contact line for a spreading droplet of initial volume $V$, we consider the following set of equations:}

%. In order to obtain the equilibrium contact angle, one needs to solve , 

%\C{(see Supplemental Material for more detail)}

%%In the intermediate range $1 \ \rm{K} < \Delta T < 11\ \rm{K}$, the spreading of the droplet ceases right after the inertial-capillary regime, as highlighted by the blue curve (experiment n$^\circ$2) in figure \ref{fig:radius}. 

%which suggests that freezing still hinders the development of the second dynamical regime. 

%\begin{figure}[b]
%\includegraphics[width=\linewidth]{fig3.pdf}% Here is how to import EPS art
%\caption{\label{fig:tic_cox_voinov} Evolution of the rescaled radius $\tilde{r}=r/R_v$ with $\tilde{t}=t/\tau_v$ in the viscous-capillary regime, for two experiments where (\textcolor{f3c1}{$\bullet$}) $T_i=-0.06$ \textcelsius\ and (\textcolor{f3c2}{$\bullet$}) $T_i=-0.64$ \textcelsius\ (in logarithmic representation). The solid lines are the associated predictions obtained by solving equations \eqref{cox_voinov}-\eqref{spherical_cap_volume}, that are extended in dotted lines prior to the observed viscous-capillary spreading. The $1/10$ slope that would correspond to a total wetting (Tanner's law) is also reported. The horizontal lines highlight the dimensionless arrest radius $\tilde{r}_m=r_m/R_v$ reached for each experiment. (b) }
%\end{figure}

\begin{align}
    \label{cox_voinov}
    \frac{d\tilde{r}}{d\tilde{t}} = \frac{1}{3 \zeta} \tan \theta(t) \left( \cos \theta_e - \cos \theta(t) \right),\\
    \label{spherical_cap_volume}
    \frac{\tilde{r}^3}{4 \sin^3 \theta (t)} \left( 2 + \cos^3 \theta(t) - 3 \cos \theta (t) \right) = 1,
\end{align}

\noindent where $\tilde{t} = t/\tau_v$, $\tilde{r} = r/R_v$ and $\tau_v = \eta R_v / \gamma$ the viscous-capillary timescale, and where $\theta(t)$ and $\theta_e$ are the instantaneous and equilibrium macroscopic contact angles, respectively. \C{Equation \eqref{cox_voinov} is obtained by balancing viscous dissipation at the contact line with the capillary forces that drive the motion (see Appendix \ref{appA} for details on the derivation), whereas relation \eqref{spherical_cap_volume} is the conservation of volume for a spherical cap \citep{2023_van_de_velde}}. In the former relation, $\zeta$ is a dimensionless parameter related to the ratio between the typical lenghtscale of the system (say, the droplet radius $R_v$) and a microscopic length of the order of a nanometer \citep{2009_bonn}. Since all the present experiments involve water droplets with a rather constant initial radius, that are deposited on the same substrate (ice), $\zeta$ can reasonably be considered constant. From there, equations \eqref{cox_voinov}-\eqref{spherical_cap_volume} can straightforwardly be solved numerically, with $\theta_e$ being evaluated from equation \eqref{spherical_cap_volume} when $r=r_m$.
\C{It should be emphasised here that $\theta_e$ is the contact angle that the drop reaches at the end of the spreading, \textit{i.e.}, at mechanical equilibrium, which is not necessarily the one defined at thermodynamic equilibrium.}

%\C{(whose derivation is provided in Supplemental Material)}

The evolution of $\tilde{r}$ with $\tilde{t}$ is illustrated in figure \ref{fig:cv_theta}(a) for two experiments where (\textcolor{f3c1}{$\bullet$}) $T_i=-0.06$ \textcelsius\ and (\textcolor{f3c2}{$\bullet$}) $T_i=-0.64$ \textcelsius. The two curves display a slow spreading rate until the droplet reaches its arrest radius $r_m$, that is highlighted by the horizontal dashed lines. The corresponding predictions from equations \eqref{cox_voinov}-\eqref{spherical_cap_volume}, with $\zeta=30$, are also reported in solid black lines in figure \ref{fig:cv_theta}(a). \C{This value for $\zeta$ has been obtained by adjusting a theoretical curve to one experiment, and has then been used to describe all tests where $\Delta T < 1$ K.} They closely match the experimental measurements for the two ice surface temperatures, from the onset of the second regime to the stabilization of the spreading drop. This result reveals that the slower spreading phase is also insensitive to phase change or to any thermal effect. The differences between the two curves, for instance in terms of the final radius, are not due to the ice temperature (as this parameter is not involved in equations \eqref{cox_voinov}-\eqref{spherical_cap_volume}) but only to fluctuations in $V$ or in the thermophysical properties of water that are captured by the present modelling. Furthermore, the dynamics is well described by a viscous-capillary evolution and the two experiments significantly depart from a 1/10 power law during the relaxation of the contact line. This key observation demonstrates that the arrest of the drop is only due to capillary effects, without any influence of temperature, and therefore that water partially wets ice. In fact, one can easily argue that thermal effects related to solidification vanish towards the capillary ones as we approach the thermodynamic equilibrium. Indeed, we can estimate the solidification timescale as $\tau_{\rm{sol}}=h^2/D_{\rm{eff}}$, where $h$ is the typical thickness of the water film at maximal spreading and $D_{\rm{eff}}$ is the effective diffusion coefficient of ice growth when water is in contact with it \citep{2024_sarlin}. Then, in the small $\Delta T$ limit, one has $D_{\rm{eff}} \propto \Delta T^2$ \C{(see more detail in Appendix \ref{appB})}. \C{On the other hand, the inertial-capillary and viscous-capillary dynamics are respectively based on the timescales $\tau_c = \sqrt{\rho {R_0}^3 / \gamma}$ and $\tau_v = \eta R_v/\gamma$ that are mostly independent of temperature.} Therefore, when comparing these two typical times \C{to the solidification timescale}, it is clear that in the limit $\Delta T \rightarrow 0$ the spreading dynamics \C{becomes unaffected by phase change as thermal equilibrium is approached, whereas for large $\Delta T$ solidification will become dominant, which agrees with the observation of the \C{partial-pinning} effect observed near the triple line for cold ice (see Appendix \ref{appB} for details about the comparison of the timescales).}

\C{Hence, for undercoolings below 1 K, the inertial-capillary dynamics is followed by a viscous-capillary regime that is well described by a model independent of thermal effects and phase change, which strongly suggests that we are closely approaching the thermodynamic limit.}

%\C{Hence, the above observations for undercoolings below 1 K, where  the inertial-capillary dynamics is followed by a viscous-capillary regime that is well described by a model independent of thermal effects and phase change, strongly suggest that we are closely approaching thermal equilibrium for those experiments.}

%strongly suggest that we are closely approaching thermal equilibrium. Indeed, equations (3.2) and  (3.3) describe well this regime, as outlined by the reviewer, with temperature and phase change playing no role in the dynamics. As this modelling compares successfully with our experiments, we infer that when $\Delta T < 1 K$ we observe a situation close enough to thermal equilibrium, at least for the duration of spreading. }

%is driven by the viscous-capillary balance towards the thermodynamical equilibrium

\subsection{The contact angle of water on ice}
\label{contact_angle}

\begin{figure}[t]
\includegraphics[width=\linewidth]{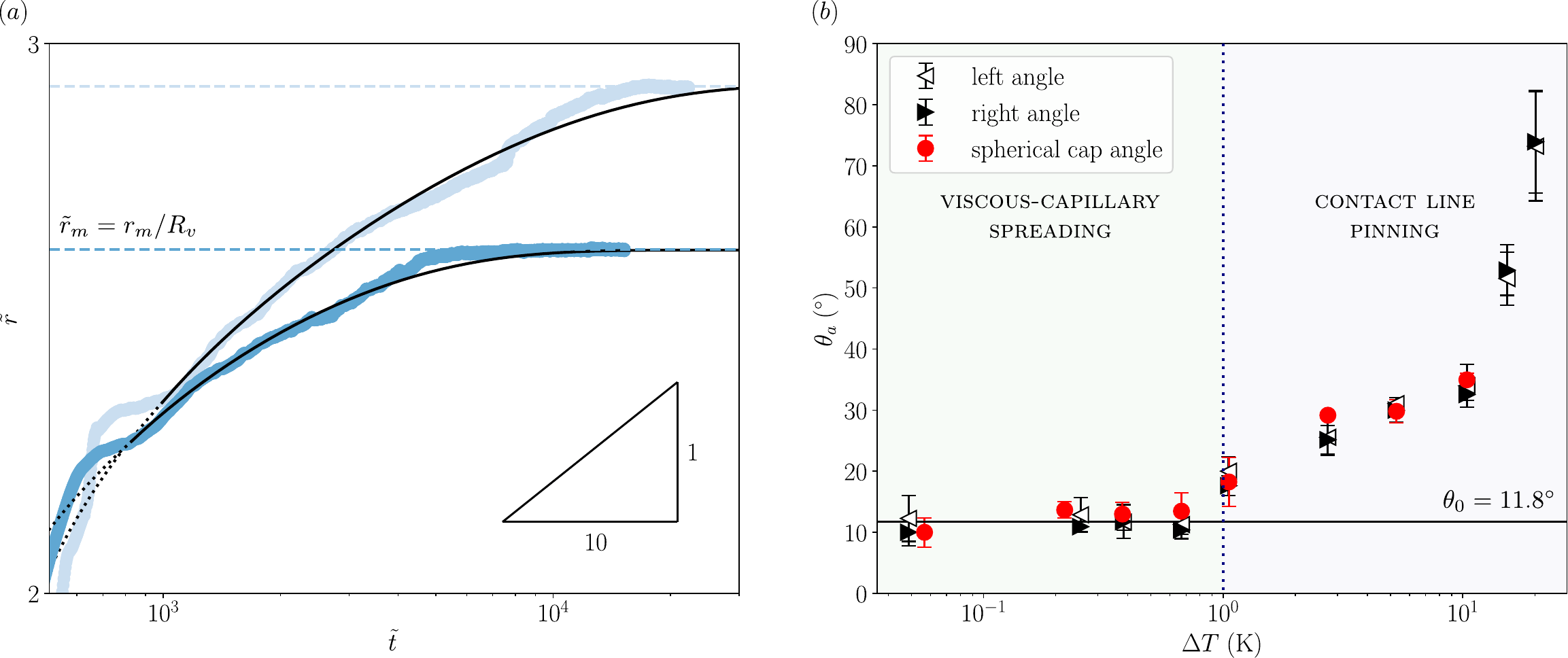}% Here is how to import EPS art
\caption{\label{fig:cv_theta} (a) Rescaled radius $\tilde{r}=r/R_v$ as a function of $\tilde{t}=t/\tau_v$, in the viscous-capillary regime, for two experiments where (\textcolor{f3c1}{$\bullet$}) $T_i=-0.06$ \textcelsius\ and (\textcolor{f3c2}{$\bullet$}) $T_i=-0.64$ \textcelsius\ (in logarithmic representation). The solid lines are the associated predictions obtained by solving equations \eqref{cox_voinov}-\eqref{spherical_cap_volume}, that are extended in dotted lines prior to the observed viscous-capillary spreading. The $1/10$ slope that would correspond to complete wetting (Tanner's law) is also indicated. The horizontal lines highlight the dimensionless arrest radius $\tilde{r}_m=r_m/R_v$ reached for each experiment. (b) Evolution of the apparent contact angle $\theta_a$ as a function of the undercooling $\Delta T$. The black and white triangles respectively correspond to measurements of the right and left angles between the water droplet and the ice once the final state is reached, using the method described by \citet{2020_quetzeri-santiago}. The red circles are the apparent angles estimated using equation \eqref{spherical_cap_volume} with $r=r_m$. The solid line is the plateau $\theta_0=11.8^\circ$, which is the mean contact angle measured from all experiments where $\Delta T < 1\mathrm{K}$.}
\end{figure}

%\C{(for more information, see Supplemental Material)}

To further discuss the value of the macroscopic equilibrium contact angle $\theta_e$, the apparent angle $\theta_a$ between the droplet and the ice layer at the end of the spreading process is systematically extracted in two different manners. Firstly, from the contours of the final photograph taken for each experiment, the right and left angles are evaluated using the algorithm developed by \citet{2020_quetzeri-santiago} to ensure robust measurements. By doing so, the contact line is approximated by first or second order polynomials, and the most reliable tangent at the triple line is retained. Secondly, an independent estimate is obtained by using equation \eqref{spherical_cap_volume} with the initial volume $V$ and the final measured radius $r_m$, assuming the final liquid film has a spherical cap shape. This method is only applied to experiments where $\Delta T \leqslant 11\ \rm{K}$, due to the reasons mentioned in section \ref{setup}. The results are illustrated in figure \ref{fig:cv_theta}(b), where $\theta_a$ is presented as a function of the undercooling $\Delta T$. The apparent angle is large when $\Delta T$ is greater than 11 \ K, as the contact line gets pinned during the short-time dynamics where its curvature is important. It then decays when the undercooling is reduced. A first inflection of the curve is observed around $\Delta T \simeq 11\ \rm{K}$, \textit{i.e.}, when contact line pinning occurs at the end of the inertial-capillary regime. Finally, $\theta_a$ reaches a constant value $\theta_0$ that is seen over more than a decade when $\Delta T < 1\ \rm{K}$, for both measurement methods. The average value and standard deviation for this constant macroscopic angle are

\begin{align}
    \label{theta_0}
    \theta_0 = 11.8^\circ \pm 1.2 ^\circ.
\end{align}

%\begin{figure}[ht]
%\includegraphics[width=\linewidth]{fig4.pdf}% Here is how to import EPS art
%\caption{\label{fig:contact_angles} Evolution of the apparent contact angle $\theta_a$ as a function of the undercooling $\Delta T$. The black and white triangles respectively correspond to measurements of the right and left angles between the water droplet and the ice once the final state is reached, using the method described by \citet{2020_quetzeri-santiago}. The red circles are the apparent angles estimated using equation \eqref{spherical_cap_volume} with $r=r_m$. The solid line is the plateau $\theta_0=11.8^\circ$, which is the mean contact angle measured from all experiments where $\Delta T < 1\mathrm{K}$.}
%\end{figure}

\noindent Interestingly, this estimate is significantly above zero, even including measurement errors from both methods. Figure \ref{fig:cv_theta}(b), combined with the aforementioned discussion on the spreading dynamics that is insensitive to thermal effects close enough to the melting point, allows us to to conclude that $\theta_0$ \textcolor{black}{is the advancing contact angle of water on ice. In other words, we demonstrated that the contact angle of water on ice exists in the thermodynamic limit and, due to the nano-smooth surface state of ice at small undercoolings, we expect $\theta_0$ to be very close to the equilibrium contact angle as contact line hysteresis must be small. This value of $\theta_0$ remains thermally-unaffected up to undercoolings of about 1 K.} Beyond this limit, thermal effects start to play a role and the apparent contact angle thereby increases. Finally, it is striking to observe that the value found for this angle agrees with some previous measurements \citep{1967_knight,2020a_thievenaz,2023_demmenie} in spite of the differences with the present work. 
%\C{Indeed, these studies were \textit{a priori} relatively far from the thermodynamic equilibrium, and these authors did not report the viscous-capillary dynamics. Thus, they were not able to assert that their measured apparent angle was the result of capillary forces only.}

\section{Conclusion}
\label{sec_conclusion}

In the present investigation, droplet deposition experiments are reported that demonstrate the partial wetting of water on ice. Indeed, close enough to the melting point of water, \textit{i.e.}, when $\Delta T < 1\ \rm{K}$, the spreading dynamics is found to be independent of thermal effects, leading to the relaxation of the liquid film to a sessile droplet shape. Within this temperature range, a short-time inertial-capillary regime takes place first and is followed by a viscous-capillary dynamics. At the end of the spreading process, a macroscopic advancing contact angle of water on ice is evidenced and is found to be very close to $12 ^\circ$. If this confirms previous estimates, the key findings here are to show that the slow evolution of the contact line in the second regime corresponds to a \C{viscous-capillary} dynamics, which implies that the observed angle is only due to wetting without any thermal effects, and that there is a smooth asymptotic convergence of the contact angle with decreasing undercooling. These two elements put together allow us to assert that \C{for $\Delta T < 1\ \rm{K}$ we significantly approach thermal and mechanical equilibria. Furthermore, the ice layer is dry due to the reduced humidity and nano-smooth, so that the contact angle hysteresis is expected to remain small. Although reducing the relative humidity implies that the experiments are not performed at vapor pressure equilibrium, our measurements agree remarkably with similar studies that operated under different relative humidity conditions \citep{1967_knight,2020a_thievenaz,2023_demmenie,2024_grivet,2025_demmenie}, thereby suggesting that this parameter does not have a significant influence. Therefore, the value of the advancing contact angle $\theta_0$ reported here is very close to the equilibrium contact angle of water on its solid phase}. 

%Indeed, the macroscopic contact angle $\theta_0$ and the values of $\gamma_{iw}$ and $\gamma$, all taken at thermodynamic equilibrium, yield $\gamma_{iv} \simeq 105 \pm 4\ \rm{mN.m^{-1}}$ using relation \eqref{eq:young}. 

\C{As mentioned in the introduction of the present work, there exist two main groups of results: a microscopic contact angle around or less than one degree reported by thermodynamic studies and a macroscopic contact angle, that is higher, arising from fluid mechanics studies \citep{2025_huerre}. While the present results show that the (macroscopic) equilibrium contact angle of water on ice is very close to $12^\circ$, it already appears that there is a clear need to understand the origin of the discrepancy between small- and large-scale observations. Solving this issue is of great interest, and deserve further investigations.}

A significant outcome of the present study is that it allows us to estimate precisely the ice-air surface energy, $\gamma_{iv}$. \C{For that purpose, one can use the Young-Dupr\'{e} relation (equation \eqref{eq:young}) to write $\gamma_{iv} = \gamma_{iw} + \gamma \cos \theta_0$, with $\theta_0=12^\circ$ and $\gamma = 75.6\ \rm{mN.m^{-1}}$. For $\gamma_{iw}$, the two extreme values available in the literature \citep{2025_huerre} are $27\ \rm{mN.m^{-1}}$ and $35\ \rm{mN.m^{-1}}$. This leads to $\gamma_{iv} \simeq 101\ \rm{mN.m^{-1}}$ or $\gamma_{iv} \simeq 109\ \rm{mN.m^{-1}}$, respectively. By taking the mean value between these two estimates, we thereby obtain $\gamma_{iv} \simeq 105 \pm 4 \ \rm{mN.m^{-1}}$. So far, several studies gave different estimates for this parameter \citep{2025_huerre}, spanning a range from 70 to 120 $\rm{mN.m^{-1}}$ that is large notably due to the lack of consensus on the contact angle of water on ice. As a result, the findings reported here, in particular the robust measurement of $\theta_0$, allow us to obtain a consistent value for $\gamma_{iv}$ that goes in the sense of \citet{1969_ketcham} and \citet{2004_pruppacher}}.

%This is a notably more accurate range that the one inferred from past studies.

%we are at thermodynamic equilibrium

We anticipate these results to be of significant importance to better understand the formation of ice structure resulting from the interaction between capillary flows and phase change. The knowledge of the equilibrium angle as well as the ice-air surface energy is also expected to be valuable in order to improve current numerical simulations dealing with three-phase problems. Given the detailed dynamical description of the spreading of a water droplet on ice provided here, this model experiment could constitute a benchmark test for the development of numerical tools. More broadly, the outcomes of the present study bring new insights on the manner water covers the ice, and could thereby be employed for the study of ice melt in glaciology.

\section*{Acknowledgments.}
The authors warmly thank C. Frot and A. Garcia for their help in the elaboration of the experimental set-up.\\
W. Sarlin and D. V. Papa contributed equally to the present study.\\
This work was supported by Agence de l'Innovation de D\'efense (AID) - via Centre Interdisciplinaire d'Etudes pour la D\'efense et la S\'ecurit\'e (CIEDS) - (project 2021 - ICING).\\
The authors report no conflict of interest.\\
The data that support the findings of this study are available from the corresponding author, upon reasonable request.\\

\section*{Appendix}
\section{The viscous-capillary spreading}
\label{appA}

Once the inertial-capillary dynamics end, a second regime takes place that involves this time a competition between surface tension, that drives the motion, and viscous dissipation. In what follows, we denote by $U$ the velocity of the contact line, $r$ the radius of the wetted area, $\theta$ the macroscopic contact angle and $u$ the velocity of the fluid. We assume that $u$ has a parabolic profile with respect to the vertical direction, so that

\begin{equation}
    \label{poiseuille}
    u(z) = \frac{3}{2} U \left( \frac{2z}{\xi} - \frac{z^2}{\xi^2} \right),
\end{equation}

\noindent with $z$ the vertical distance from the substrate and $\xi=\alpha \tan \theta$ the position of the free-surface, that depends on the horizontal distance $\alpha$ from the triple point. In that situation, the power density is $\eta (\partial u/\partial z)^2$ \citep{2012_guyon}, so that the dissipation per unit length reads

\begin{equation}
    \label{dissipation}
    D = \int_{\alpha_m}^{\alpha_M} \int_0^\xi \eta \left( \frac{\partial u}{\partial z} \right)^2 \,dz d\alpha = \frac{3 \eta U^2}{\tan \theta} \zeta,
\end{equation}

\noindent where $\zeta=\ln \left( \alpha_M/\alpha_m \right)$ depends on two quantities $\alpha_m$ and $\alpha_M$, of the order of a nanometer and of the typical size of the system, respectively \citep{2009_bonn}. 

On the other hand, the capillary force per unit length which acts on the contact line can be expressed as $C=\gamma_{sg}-\gamma \cos \theta - \gamma_{sl}$, with $\gamma_{sg}$ and $\gamma_{sl}$ the surface tensions associated with the solid-gas and solid-liquid interfaces, respectively. As $C$ vanishes when the equilibrium contact angle $\theta_e$ is reached, we also have $\gamma_{sg}=\gamma_{sl}+\gamma \cos \theta_e$. Thus, $C$ reads $C = \gamma \left( \cos \theta_e - \cos \theta \right)$ and the power (per unit length) this force induces is $CU$. If it is fully dissipated, we thereby have

\begin{equation}
    \label{viscous_capillary_balance_1}
    \gamma \left( \cos \theta_e - \cos \theta \right) U = \frac{3 \eta \zeta}{\tan \theta} U^2 ,
\end{equation}

\noindent which can be rewritten, as $U=dr/dt$, in the form

\begin{equation}
    \label{viscous_capillary_balance_2}
    \frac{d \tilde{r}}{d \tilde{t}} = \frac{1}{3 \zeta} \tan \theta \left( \cos \theta_e - \cos \theta \right),
\end{equation}

\noindent with $\tilde{r} = r / R_v$ and $\tilde{t} = t / \tau_v$ ($R_v = (3V/(4 \pi))^{1/3}$ being the equivalent radius based on the initial volume $V$ of liquid deposited, and $\tau_v = \eta R_v / \gamma$ the typical viscous-capillary timescale). Equation \eqref{viscous_capillary_balance_2} is the same as equation \eqref{cox_voinov} in the main text.

A case of particular interest is the situation of complete wetting, \textit{i.e.}, when $\theta_e = 0$. When this is the case, assuming $\theta \ll 1$, equation \eqref{viscous_capillary_balance_2} reduces to

\begin{equation}
    \label{viscous_capillary_balance_3}
    \frac{d r}{d t} = \frac{\gamma}{6 \eta \zeta} \theta^3 \propto \frac{\gamma}{\eta} \theta^3,
\end{equation}

\noindent as $\zeta$ can approximately be considered constant \citep{1984_de_gennes}. Then, by observing that, for small angles, we have $\theta \simeq 2 e / r$ for a spherical cap of radius $r$ and uppermost height $e$, and that a rough conservation of volume implies $r^2e \propto {R_v}^3$, equation \eqref{viscous_capillary_balance_3} leads to

\begin{equation}
    \label{tanner_diff_equ}
    r^9 \frac{d r}{d t} \propto \frac{\gamma}{\eta} {R_v}^9,
\end{equation}

\noindent which, after integration, gives

\begin{equation}
    \label{tanner_law}
    \frac{r}{R_v} \propto \left( \frac{t}{\tau_v} \right)^{1/10}.
\end{equation}

\noindent Equation \eqref{tanner_law} is widely known as Tanner's law of spreading for a viscous drop made of a completely wetting liquid \citep{1979_tanner}. %Conversely, the expression \eqref{viscous_capillary_balance_2} is also known as the Cox-Voinov law of spreading for a partially wetting fluid \cite{1986_cox,1976_voinov}.

%\section{The Cox-Voinov relaxation}
%
%\begin{figure*}[t]
%\includegraphics[width=\linewidth]{SM_fig2.pdf}% Here is how to import EPS art
%\caption{\label{fig:sm_fig2} Evolution of $\tilde{r}$ as a function of $\tilde{t}$ in the viscous-capillary regime, as obtained when solving equations \eqref{cox_voinov} and \eqref{spherical_cap_volume} for different values of (a) $\zeta$ and (b) $\theta_e$. Here, the liquid volume $V$ is $2.15 \times 10^{-8}\ \rm{m^3}$ and the surface tension is $\gamma \simeq 76\ \rm{mN.m^{-1}}$, which corresponds to the values measured for the experiment where (\textcolor{f3c1}{$\bullet$}) $T_i=-0.06$ \textcelsius\ in the main text.}
%\end{figure*}

\section{Spreading and solidification timescales}
\label{appB}

The time needed to solidify a droplet of volume $V$ and radius $R_v = (3V/(4 \pi))^{1/3}$ can be estimated via the self-similar solution of the Stefan problem, that gives in particular the evolution of the solidification front $h = \sqrt{D_{\rm{eff}}t}$. In this expression, $D_{\rm{eff}}$ is an effective diffusion coefficient that is solution to a transcendental equation involving the thermal properties of the different phases and the undercooling $\Delta T$ \citep{2019_thievenaz,2024_sarlin}. Considering that, at maximal spreading $r_m$, the droplet has a characteristic height $h \propto {R_v}^3/{r_m}^2$, the typical solidification timescale writes

\begin{equation}
    \label{eq:tau_sol}
    \tau_{\rm{sol}} = \frac{{R_v}^6}{{r_m}^4 D_{\rm{eff}}}.
\end{equation}

\noindent Similarly, it is possible to estimate the timescale $\tau_{\rm{ic}}$ needed to reach the maximal spreading diameter in the inertial-capillary regime from the scaling law $r/R_0 \propto (t/\tau_c)^{1/2}$, that yields (here, we take $R_0 \simeq R_v$ for the sake of simplicity)

\begin{equation}
    \label{eq:tau_ic}
    \tau_{\rm{ic}} = \sqrt{\frac{\rho {R_v}^3}{\gamma}} \left( \frac{r_m}{R_v} \right)^2.
\end{equation}

\noindent Conversely, the viscous-capillary timescale $\tau_{\rm{vc}}$ is obtained from the scaling $r/R_v \propto (\gamma t/(\eta R_v))^{1/10}$ (see Appendix \ref{appA}), that leads this time to

\begin{equation}
    \label{eq:tau_vc}
    \tau_{\rm{vc}} = \frac{\eta r_v}{\gamma} \left(\frac{r_m}{R_v} \right)^{10}.
\end{equation}

With these three times, it is possible to derive two dimensionless ratios comparing the solidification timescale to the inertial-capillary or viscous-capillary ones. They respectively write

\begin{gather}
\label{eq:tau_ratios}
    \frac{\tau_{\rm{sol}}}{\tau_{\rm{ic}}} = \frac{1}{D_{\rm{eff}}} \sqrt{\frac{\gamma R_v}{\rho }} \left(\frac{R_v}{r_m} \right)^{6} \qquad \mathrm{and} \qquad \frac{\tau_{\rm{sol}}}{\tau_{\rm{vc}}} = \frac{1}{D_{\rm{eff}}} \frac{\gamma R_v}{\eta} \left( \frac{R_v}{r_m} \right)^{14}.
\end{gather}

\noindent In both cases, one can observe that they are inversely proportional to $D_{\rm{eff}}$. Then, for undercoolings approaching 0, one has 

\begin{equation}
    \label{eq:Deff0}
    D_{\rm{eff}} \sim \frac{4 D_i {c_{p,i}}^2}{\pi {\mathcal{L}_f}^2} \Delta T^2,
\end{equation}

\noindent with $D_i$ and $c_{p,i}$ the ice thermal diffusivity and heat capacity, and $\mathcal{L}_f$ the latent heat of fusion for water \citep{2019_thievenaz}. As a result, when $\Delta T \to 0$ the ratios of timescales will tend to infinity as fast as $\Delta T^2$. This implies that, close to the melting point, the solidification rate rapidly becomes negligible compared to the spreading dynamics, regardless of the drop size or the maximal spreading radius. On the other hand, for large undercoolings the solidification rate becomes important and causes the partial-pinning effect observed near the triple line.

\bibliographystyle{jfm}
\bibliography{bibliography}

%Use of the above commands will create a bibliography using the .bib file. Shown below is a bibliography built from individual items.

%% End of file `jfm.bib'.

\end{document}